\documentclass[conference]{IEEEtran}
\IEEEoverridecommandlockouts
\usepackage{graphicx}
\usepackage{hyperref}
\usepackage{balance}
\usepackage{amsmath}
\usepackage{amssymb}
\usepackage{multicol}

\usepackage{stfloats}

\usepackage{epsf}
\usepackage[usenames]{color}

\begin{document}
\hyphenation{multi-symbol}
\title{Optimization of a Finite Frequency-Hopping \\ Ad Hoc Network in Nakagami Fading \vspace{-0.2cm}}
\author{\IEEEauthorblockN{Matthew C. Valenti,\IEEEauthorrefmark{1}
Don Torrieri,\IEEEauthorrefmark{2}
and Salvatore Talarico\IEEEauthorrefmark{1} }
\IEEEauthorblockA{\IEEEauthorrefmark{1}West Virginia University, Morgantown, WV, USA. \\
\IEEEauthorrefmark{2}U.S. Army Research Laboratory, Adelphi, MD, USA.}
\vspace{-0.5cm}
\thanks{The authors were sponsored by the National Science Foundation under Award No. CNS-0750821 and by the United States Army Research Laboratory under Contract W911NF-10-0109.}
}
\date{}
\maketitle

\thispagestyle{empty}

\begin{abstract}
This paper considers the analysis and optimization of a frequency-hopping ad hoc network with a finite number of mobiles and finite spatial extent.  The mobiles communicate using coded continuous-phase frequency-shift keying (CPFSK) modulation.   The performance of the system is a function of the number of hopping channels, the rate of the error-correction code, and the modulation index used by the CPFSK modulation.   For a given channel model and density of mobiles, these parameters are jointly optimized by maximizing the (modulation-constrained) transmission capacity, which is a measure of the spatial spectral efficiency of the system.  The transmission capacity of the finite network is found by using a recent expression for the spatially averaged outage probability in the presence of Nakagami fading, which is found in closed form in the absence of shadowing and can be solved using numerical integration in the presence of shadowing.
\end{abstract}

\section{Introduction} \label{Section:Intro}
An \textit{ad hoc network} or peer-to-peer network comprises autonomous systems that communicate without a centralized control or assistance. Ad hoc networks include mobile communication networks that possess no supporting infrastructure, fixed or mobile; usually each mobile has identical signal processing capability.
The preferred channel access for ad hoc networks is direct-sequence or frequency-hopping (FH) spread spectrum.  This paper focuses specifically on frequency-hopping spread spectrum ad hoc networks.   Such networks are characterized by independent, identical, FH radios that share the same carriers and frequency channels, and are nearly stationary in location over a single hop duration.

The use of spread spectrum in ad hoc networks provides the many advantages of code-division multiple-access communications \cite{andrews:2007}.  Two major advantages of FH are that it can be implemented over a much larger frequency band than  is possible with direct-sequence spreading, and that the band can be divided into noncontiguous segments. Another major advantage is that FH provides resistance to multiple-access interference while not requiring power control to prevent the near-far problem. Since direct-sequence and narrowband systems cannot escape the near-far problem by hopping, accurate power control is crucial in limiting the impact of the near-far problem in cellular networks \cite{torrieri:2011}. However, power control is not viable for ad hoc networks because of the lack of a centralized architecture.  Therefore, direct-sequence networks require a guard zone imposed by some type of carrier-sense multiple access with collision detection \cite{torrieri:2012a}. In contrast, FH networks can use an ALOHA-type multiple-access protocol.

%In the quest for analytical tractability, the recent literature (e.g., \cite{andrews:2010}, \cite{win:2009}) has analyzed ad hoc networks under the assumption of an infinite number of mobiles spread over an infinite area. The locations of the mobiles at any time instant are assumed to be distributed as a Poisson point process. When the network is finite, the Poisson point process is a poor model because it allows an unbounded number of mobiles. The model is also unsatisfactory when the transmissions are coordinated by an interference-avoidance protocol, which imposes a minimum separation between interfering transmitters. However, this model is extremely useful because it enables authors to use Campbell's theorem \cite{stoyan:1996}, which leads to tractable mathematical expressions that vastly simplify the performance analysis.

For FH systems, continuous-phase frequency-shift keying (CPFSK) is the preferred modulation.   Frequency hopping with CPFSK offers a constant-envelope signal, a compact signal spectrum, and robustness against both partial-band and multiple-access interference \cite{torrieri:2011}.  CPFSK modulation is characterized by its modulation order, which is the number of possible tones, and by its modulation index $h$, which is the normalized tone spacing (assumed here to be constant).   For a fixed modulation order, the selection of $h$ involves a tradeoff between bandwidth and performance \cite{cheng:ciss2007}.  The resistance to frequency-hopping multiple-access interference generally increases with decreasing $h$ while the error-rate generally decreases with increasing $h$.
%In an ad hoc network, the tradeoff can be quantified by establishing a threshold on the signal-to-interference-and-noise ratio (SINR), which quantifies the minimum signal quality required for reliable communication.   In \cite{cheng:ciss2007}, we computed the capacity of noncoherently detected CPFSK, and in \cite{torrieri:2008} we optimized a FH system with a fixed number of hopping frequencies with respect to this capacity.
When the ad hoc network uses channel-coded CPFSK modulation, the performance is a function of not only the modulation index, but also the rate of the error-control code and the number of frequency-hopping channels.   As each of these parameters influences bandwidth and energy efficiency, there is a tradeoff among them, and the main goal of this paper is to gain some insight into this tradeoff.

An appropriate objective for the optimization of the network's operational parameters is to maximize the {\em transmission capacity} \cite{weber:2005,weber:2010}, which is a measure of the spatial spectral efficiency and is constrained in this paper to explicitly take into account the CPFSK modulation format.   In contrast with the recent literature (e.g., \cite{andrews:2010}, \cite{win:2009}), which assumes an infinite number of mobiles spread over an infinite area, in this paper the spatial extent of the network and number of mobiles are finite.   Each mobile has a uniform location distribution with an allowance for the mobile's duty factor and shadowing.  The analysis uses a recent expression \cite{torrieri:2012} for the exact outage probability in the presence of Nakagami fading conditioned on the network topology, which is determined by both the mobile locations and the shadowing factors. The modulation-constrained transmission capacity is found by averaging the outage probability over the network topoloty, which is possible in closed form in certain cases (i.e., no shadowing, and interfering mobiles uniformly distributed within an annulus centered at the reference receiver) and through numerical integration in other cases (e.g., networks with shadowing).   By maximizing the modulation-constrained transmission capacity, the optimal set of operating parameters (code rate, modulation index, and number of frequency channels) is found for a given channel model and density of mobiles.

\pagebreak

The remainder of the paper is organized as follows.  Section \ref{Section:SystemModel} presents a system model, which culminates in an expression for the instantaneous SINR at the reference receiver.  Section \ref{Section:Outage} reviews a new expression \cite{torrieri:2012} for the conditional outage probability; i.e., the probability that the SINR is below a threshold, given a particular set of mobile locations and shadowing factors.  Section \ref{Section:AverageOutage} discusses how to remove the conditioning by averaging over the spatial and shadowing distributions.   Section \ref{Section:TC} discusses the modulation-constrained transmission capacity, which is the area spectral efficiency of the network and the metric used for the optimization.  Section \ref{Section:Optimization} describes the procedure used for determining the combination of parameters that maximize the transmission capacity and gives optimization results for several channel and network models.  %Finally, the paper concludes in Section \ref{Section:Conclusion}.

\section{Network Model} \label{Section:SystemModel}
The network comprises $M+2$ mobiles that include
a receiver, a reference (source) transmitter $X_{0}$, and $M$ interfering
transmitters $X_{1},...,X_{M}.$ The coordinate system is selected such that
the receiver is at the origin.   The variable $X_{i}$ represents
both the $i^{th}$ transmitter and its location, and $||X_{i}||$ is the
distance from $X_i$ to the receiver.    While the interferers can be located in any arbitrary region, we assume they are located in an annular region with inner radius $r_{\mathsf{ex}}$ and outer radius $r_{\mathsf{net}}$.  A nonzero $r_{\mathsf{ex}}$ may be used to model the effects of interference-avoidance protocols \cite{torrieri:2012a}.

\setcounter{equation}{14}
\begin{figure*}[!b]
\vspace{0cm}
\hrulefill
\begin{eqnarray}
\bar{F}_{\mathsf Z}(z) \hspace{-0.25cm}
&=&
\hspace{-0.25cm}
e^{-\beta _{0}z}
\hspace{-0.15cm}
\sum_{s=0}^{m_{0}-1}{\left( \beta
_{0}z\right) }^{s}\sum_{t=0}^{s}\frac{z^{-t}}{(s-t)!}
\hspace{-0.35cm}
\mathop{ \sum_{\ell_i \geq 0}}_{\sum_{i=0}^{M}\ell _{i}=t}
\hspace{-0.15cm}
\prod_{i=1}^{M}%
\left[  (1-p_{i})\delta _{\ell _{i}}+\frac{2p_{i}\Gamma
(\ell _{i}+m_{i})m_{i}^{m_{i}}\left[
J\left( c_{i}r_{\mathsf{net}}^{\alpha }\right) -J\left( c_{i}r_{\mathsf{ex}}^{\alpha }\right) %
\right]}{\alpha
c_{i}^{2/\alpha }(r_{\mathsf{net}}^{2}-r_{\mathsf{ex}}^{2})(\ell _{i}!)\Gamma (m_{i})\beta
_{0}^{(m_{i}+\ell _{i})}\left( m_{i}+\frac{2}{\alpha }\right) } \right]
\label{cdf} \hspace{+0.35cm}
\end{eqnarray}
\vspace{-0.75cm}
\end{figure*}
\setcounter{equation}{0}

$X_{i}$ transmits a signal whose average received power in the absence of fading and
shadowing is $P_{i}$ at a reference distance $d_{0}$.
At the receiver, $X_i$'s  power is
\begin{eqnarray}
  \rho_i
  & = &
  P_i g_i 10^{\xi_i/10} f( ||X_i|| ) \label{eqn:power}
\end{eqnarray}
where  $g_i$ is the power gain due to fading, $\xi_i$ is a shadowing coefficient, and $f( ||X_i|| )$ is a path-loss function.  Each $g_i = a_i^2$, where $a_i$ is Nakagami with parameter $m_i$, and $\mathbb E[g_i] = 1$.  In Rayleigh fading, $m_i=1$ and $g_i$ is exponential.  In the presence of log-normal shadowing, the $\{ \xi_i \}$ are i.i.d. zero-mean Gaussian with standard deviation $\sigma_s$ dB.  In the absence of shadowing, $\xi_i = 0$.  \ For $d\geq d_{0}$, the path-loss function is
expressed as the attenuation power law
\begin{eqnarray}
   f \left( d \right)
   & = &
   \left( \frac{d}{d_0} \right)^{-\alpha} \label{eqn:pathloss}
\end{eqnarray}
where $\alpha > 2$ is the attenuation power-law exponent, and $d_0$ is sufficiently large that the signals are in the far field.

Channel access is through a synchronous frequency-hopping protocol.  The hopping is slow, with multiple symbols per hop, which is a more suitable strategy for ad hoc networks than fast hopping \cite{torrieri:2011}.  An overall frequency band of $B$ Hz is divided into $L$ frequency channels, each of bandwidth $B/L$ Hz.  The transmitters independently select their transmit frequencies with equal probability.  Let $p_i$ denote the probability that interferer $X_i$ uses the same frequency as the source.  Let
$D_i \leq 1$ be the duty factor of the interferer.  It follows that $p_i=D_i/L$ and that using a duty factor less than unity is equivalent to hopping over more than $L$ frequencies \cite{torrieri:2011}.  Assuming that $D_i=D$ for all interferers, $L'=L/D$ denotes the {\em equivalent} number of frequency channels.  It is assumed that the \{$g_{i}\}$ remain fixed for the duration of a hop, but vary independently from hop to hop (block fading).  While the $\{g_{i}\}$ are independent, they are not necessarily identically distributed.  The channel from each transmitting mobile to the reference receiver can have a distinct Nakagami parameter $m_i$.

The instantaneous SINR at the receiver is
\begin{eqnarray}
   \gamma
   & = &
   \frac{ \rho_0 }{ \displaystyle {\mathcal N} + \sum_{i=1}^{M} I_i \rho_i } \label{Equation:SINR1}
\end{eqnarray}
where $\mathcal N$ is the noise power and $I_i$ is a variable that indicates the presence and type of interference (i.e. co-channel interference or adjacent-channel interference).   When adjacent-channel interference is neglected, $I_i=1$ when $X_i$ selects the same frequency as $X_0$, and $I_i=0$ otherwise.  It follows that $I_i$ is Bernoulli with probability $P[I_i=1]=p_i$.

% Adjacent-channel interference due to spectral splatter \cite{torrieri:2011} can be taken into account by setting $I_i$ equal to the fraction of in-band power (typically 99 percent) when $X_i$ selects the same frequency as $X_0$, and equal to the fraction of power in the adjacent channel when $X_i$ selects a frequency adjacent to the one selected by $X_0$ (typically 0.5 percent).

% This is a {\em physical interference model} \cite{cardieri:2010}.

Substituting (\ref{eqn:power}) and (\ref{eqn:pathloss}) into (\ref{Equation:SINR1}),  the SINR is
%   \gamma
%   & = &
%   \frac{ P_0 g_0 10^{\xi_0/10} }{ \displaystyle {\mathcal N} + \sum_{i=1}^M P_i I_i g_i 10^{\xi_i/10} ||X_i||^{-\alpha}}
%   \nonumber \\
%   & = &
\begin{eqnarray}
   \gamma
   & = &
   \frac{ g_0 \Omega_0  }{ \displaystyle \Gamma^{-1} + \sum_{i=1}^M I_i g_i \Omega_i }
   \label{Equation:SINR2}
\end{eqnarray}
where $\Gamma = r_0^\alpha P_{0}/\mathcal{N}$ is the signal-to-noise
ratio (SNR)  when the transmitter is at unit distance and fading and
shadowing are absent, $\Omega_i = (P_i/P_0)10^{\xi_i/10} ||X_i||^{-\alpha}$ is the normalized power of $X_i$, and $\Omega_0 = 10^{\xi_0/10}||X_0||^{-\alpha}$.

\section{Conditional Outage Probability} \label{Section:Outage}
\label{Section:OutageProbability}
Let $\beta$ denote the minimum SINR required for reliable reception and $\boldsymbol{\Omega }=\{\Omega_{0},...,\Omega _{M}\}$ represent the set of normalized powers.  An \emph{outage} occurs when the SINR falls below $\beta$.  Conditioning on $\boldsymbol{\Omega }$, the outage probability is
\begin{eqnarray}
   \epsilon_o
   & = &
   P \left[ \gamma \leq \beta \big| \boldsymbol \Omega \right].
   \label{Equation:Outage1}
\end{eqnarray}
Because it is conditioned on $\boldsymbol{\Omega }$, the outage probability depends on the locations of the mobiles and the shadowing factors, which have dynamics over timescales that are much slower than the fading.
By defining a variable
\vspace{-0.25cm}
\begin{eqnarray}
 \mathsf Z & = & \beta^{-1} g_0 \Omega_0 - \sum_{i=1}^M I_i g_i \Omega_i \label{eqn:z}
\end{eqnarray}
the conditional outage probability may be expressed as
\begin{eqnarray}
  \epsilon_o
  & = &
  P
  \left[
   \mathsf Z  \leq \Gamma^{-1} \big| \boldsymbol \Omega
  \right]
  = F_{\mathsf Z} \left( \Gamma^{-1} \big| \boldsymbol \Omega \right) \label{Equation:OutageCDF}
\end{eqnarray}
which is the cumulative distribution function (cdf) of $\mathsf Z$ conditioned on $\boldsymbol \Omega$ and evaluated at $\Gamma^{-1}$.

 Under the restriction that the Nakagami parameter $m_0$ of the source's channel is integer-valued, the complementary cdf of $\mathsf{Z}$ conditioned on $\boldsymbol{\Omega}$ is shown in \cite{torrieri:2012} to be
\begin{eqnarray}
\bar{F}_{\mathsf Z}\left( z \big| \boldsymbol \Omega \right)
& = &
e^{-\beta_0 z }
\sum_{j=0}^{m_0-1} {\left( \beta_0 z \right)}^j
 \sum_{k=0}^j
\frac{ z^{-k} H_k ( \boldsymbol \Psi )}{ (j-k)! }
\label{Equation:NakagamiConditional}
\end{eqnarray}
where $\beta_0 = m_0 \beta/\Omega_0$,
\begin{eqnarray}
   \Psi_i
   & = &
   \left(
      \beta_0 \frac{\Omega_i}{m_i} + 1
    \right)^{-1}\hspace{-0.5cm}    \label{Equation:Psi}\\
   H_k ( \boldsymbol \Psi )
   & = &
   \mathop{ \sum_{\ell_i \geq 0}}_{\sum_{i=0}^{M}\ell_i=k}
   \prod_{i=1}^M
   G_{\ell_i} ( \Psi_i ), \label{Equation:Hfunc}
\end{eqnarray}
the summation in (\ref{Equation:Hfunc}) is over all sets of positive indices that sum to $k$, and
\begin{eqnarray}
 G_\ell( \Psi_i )
 & = &
 \begin{cases}
 1 - p_i (1-\Psi_i^{m_i})  & \mbox{for $\ell=0$} \\
 \frac{
 p_i \Gamma( \ell + m_i ) }
 {\ell! \Gamma(m_i) }
 \left( \frac{\Omega_i}{m_i} \right)^{\ell} \Psi_i^{m_i+\ell} & \mbox{for $\ell>0$.} \\
 \end{cases} \label{Equation:Gfunc}
\end{eqnarray}

\vspace{-0.05cm}
\subsection{Examples}
\begin{figure}[t]
\centering
% \vspace{-0.1cm}
%\hspace{-0.5cm}
\includegraphics[width=9.25cm]{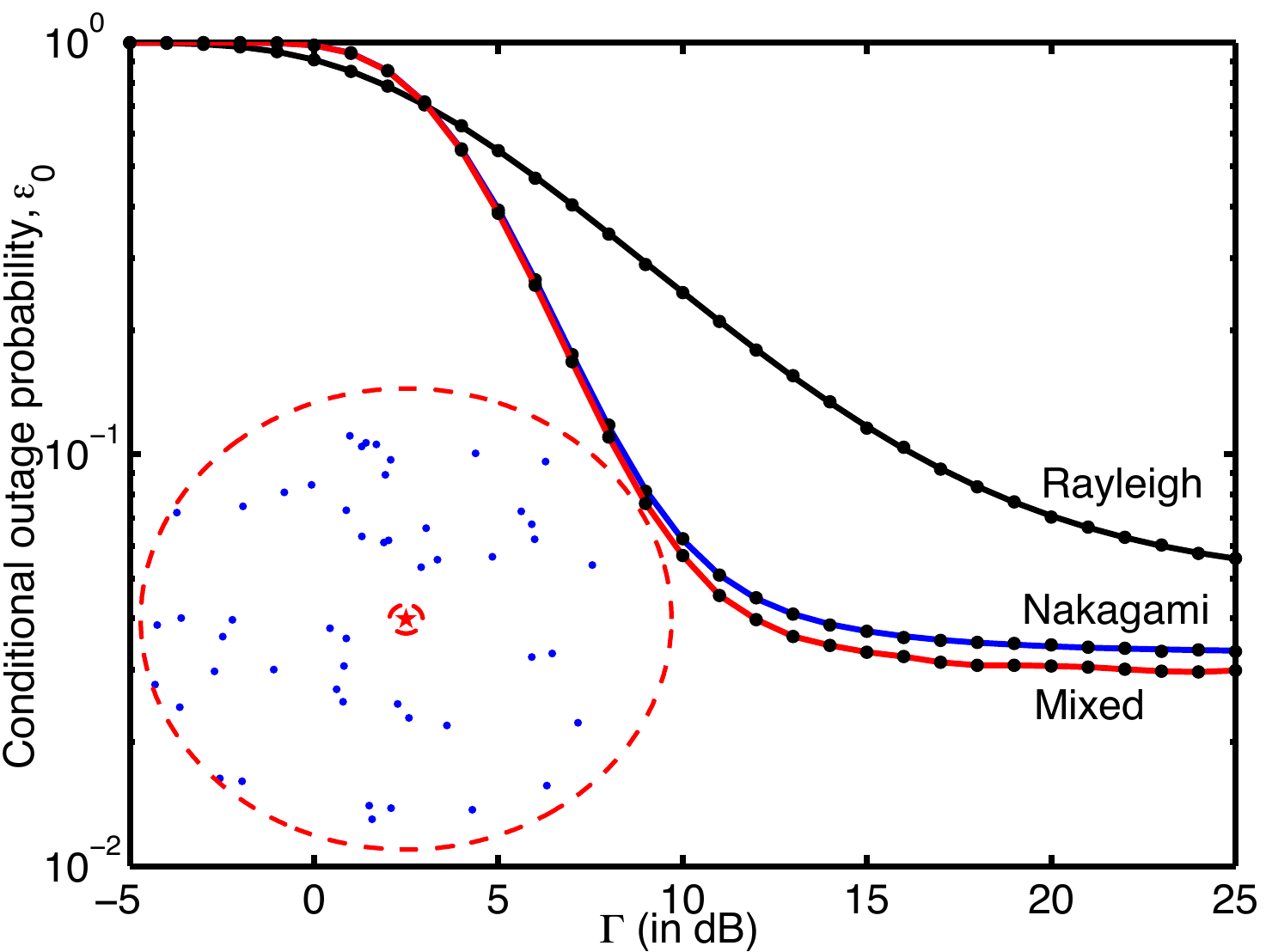}
\vspace{-0.7cm}
\caption{Conditional outage probability $\epsilon_{o}$ of Examples \#1 and \#2 as a function of SNR $\Gamma$.  Analytical curves are solid while dots represent simulated values.  Top curve: all channels Rayleigh.  Middle curve: all channels $m=4$.  Bottom curve: $m=4$ from source, and $m=1$ from interferers. The network topology is shown in the inset.  The receiver is represented by the star at the center of the radius-$4$ circle while the 50 interferers are shown as dots.  \label{Figure:Example1} }
\vspace{-0.5cm}
\end{figure}

\vspace{-0.05cm}
In the following examples, the source transmitter was placed at unit distance from the receiver; i.e., $||X_{0}|| = 1$, and fifty interferers were independently placed according to a uniform distribution in an annular region with inner dimension $r_{\mathsf{ex}} = 0.25$ and outer dimension  $r_{\mathsf{net}}=4$.  The resulting network is shown in the inset of Fig. \ref{Figure:Example1}.  The $\boldsymbol \Omega$ was determined by assuming a path-loss exponent $\alpha = 3$, a common transmit power $P_i=P_0$, and the absence of shadowing.  The equivalent number of frequency channels was set to $L'=200$, and the SINR threshold set to $\beta = 3.7$ dB.
% corresponding to the AWGN capacity limit for rate-1/2 noncoherent orthogonal binary CPFSK.

{\bf Example \#1.}  Suppose that all signals undergo Rayleigh fading.  Then $m_i=1$ for all $i$, and (\ref{Equation:NakagamiConditional})-(\ref{Equation:Gfunc}) specialize to
\vspace{-0.2cm}
\begin{eqnarray}
   \bar{F}_{\mathsf Z}(z|\boldsymbol \Omega)
   & = &
   e^{-\beta_0 z } \prod_{i=1 }^{M}   \frac {1 + \beta_0(1-p_i)\Omega_i} { 1 + \beta_0 \Omega_i}.
   \label{Equation:RayleighConditional}
\end{eqnarray}
\vspace{-0.25cm}

\noindent The outage probability was found by evaluating (\ref{Equation:RayleighConditional}) at $z=\Gamma^{-1}$ and is shown in Fig. \ref{Figure:Example1}. Also shown is a curve generated by simulation, which involved randomly generating the exponentially-distributed  $\{g_i\}$ and drawing each interferer's hopping frequency from a uniform distribution.  The analytical and simulation results coincide, which is to be expected because  (\ref{Equation:RayleighConditional}) is exact.  Any discrepancy between the curves can be attributed to the finite number of Monte Carlo trials (one million trials were executed per SNR point).

{\bf Example \#2.} Now suppose that the link between the source and receiver undergoes Nakagami fading with parameter $m_0=4$.  In this case, (\ref{Equation:NakagamiConditional})-(\ref{Equation:Gfunc}) must be used to find the conditional outage probability.  Using the same $\beta$, $L'$, and $\boldsymbol \Omega$ as Example \#1, the outage probability was found and plotted in Fig. \ref{Figure:Example1}.  Two cases were considered for the interferer's Nakagami coefficient: $m_i=1$ and $m_i=4$.  The $m_i=4$ case represents the situation where the source and interferer are equally visible to the receiver, while the $m_i=1$ case represents a more typical situation where the interferers are not in the line-of-sight. As with the previous example, the analytical curves are verified by simulations involving one million Monte Carlo trials per SNR point.

\section{Spatially Averaged Outage Probability}\label{Section:AverageOutage}

\setcounter{equation}{21}
\begin{figure*}[b!]
\vspace{0cm}
\hrulefill
\begin{eqnarray}
\bar{F}_{{\mathsf Z}_M}(z) \hspace{-0.25cm} &=&
\hspace{-0.25cm}
\sum_{s=0}^{m_0-1}
\hspace{-0.10cm}
 \sum_{t=0}^s
\frac{ z^{-t} }{ (s-t)! } \hspace{-0.35cm} \mathop{ \sum_{\ell_i \geq 0}}_{\sum_{i=0}^{M}\ell_i=t}\hspace{-0.35cm} \int_{0 }^{\infty } \exp \left\{-\frac{\beta m_0  z}{y} \right\} {\left( \frac{\beta m_0 z}{y} \right)}^s \prod_{i=1}^M \left[(1-p_i)\delta_{\ell_i}
 +
 \hspace{-0.15cm} \int_{0 }^{\infty } \hspace{-0.15cm} \kappa \left(\omega, \frac{\beta m_0 }{y}\right) \varphi(\omega)
  d \omega
\right]  f_{\Omega_0}(y) d y. \nonumber \\
\label{shad05}
\end{eqnarray}
\vspace{-0.55cm}
\end{figure*}
\setcounter{equation}{12}

Because it is conditioned on ${\boldsymbol \Omega}$, the outage probability $\epsilon_o$ presented in the last section depends on the location of the interferers and the values of the shadowing factors.  The conditioning on ${\boldsymbol \Omega}$ can be removed by averaging ${F}_{\mathsf Z}(z|\boldsymbol \Omega)$ with respect to
$\boldsymbol \Omega$, which results in the average outage probability
\begin{eqnarray}
   \epsilon
   &=&
   E_{\boldsymbol \Omega} \left[
   \epsilon_o
   \right]
   =
   E \left[
   {F}_{\mathsf Z}\left( \Gamma^{-1} \big| \boldsymbol \Omega \right)
   \right]
   =
   {F}_{\mathsf Z}\left( \Gamma^{-1} \right).
\end{eqnarray}
Finding $\bar{F}_{\mathsf Z}\left( z \right) = 1 - {F}_{\mathsf Z}\left( z \right)$ requires the following integration:
\vspace{-0.5cm}
\begin{eqnarray}
\bar{F}_{\mathsf Z}(z)
& = &
\int
\left( \prod_{i=0}^M f_{\Omega_i}( \omega_i ) \right)
\bar{F}_{\mathsf Z}(z \big| \boldsymbol \omega )  d\boldsymbol \omega
\label{cdf_M_rho}
\end{eqnarray}
where $f_{\Omega_i}(\omega_i)$ is the pdf of $\Omega_i$ and the $\{\Omega_i\}$ are assumed to be independent.

In \cite{torrieri:2012}, it is shown that $\bar{F}_{\mathsf Z}(z)$ in the absence of shadowing is given by (\ref{cdf}) at the bottom of the page,
where $c_{i}=P_{0}/P_{i}$,  % CORRECTION
\setcounter{equation}{15}
\begin{equation}
J(y)= {_{2}F_{1}\left( \left[ m_{i}\hspace{-0.05cm}+\hspace{-0.05cm}\ell _{i},m_{i}\hspace{-0.05cm}+\hspace{-0.05cm}\frac{2}{\alpha }\right]
;m_{i}\hspace{-0.05cm}+\hspace{-0.05cm}\frac{2}{\alpha }\hspace{-0.05cm}+\hspace{-0.05cm}1;-\frac{m_{i}y}{\beta _{0}}\right)}y^{m_i+\frac{2}{\alpha}}
\end{equation}
and
$_{2}F_{1}$ is  the Gauss hypergeometric function,
which has the integral representation
\begin{eqnarray}
_{2}F_{1}([a,b];c;x) & = & \nonumber \\
&  & \hspace{-2cm}
\frac{\Gamma (c)}{\Gamma (b)\Gamma (c-b)}%
\int_{0}^{1}\nu ^{b-1}(1-\nu )^{c-b-1}(1-x\nu )^{-a}d\nu.  \nonumber \label{HYPERG} \\
\end{eqnarray}%
While the hypergeometric function is itself an integral, it is widely known and is implemented as a single function call in most mathematical programming languages, including Matlab.

When shadowing is present, $\Omega_i = c_i^{-1} 10^{\xi_i/10} ||X_i||^{-\alpha}$, $i \geq 1$, which has pdf
\vspace{-0.15cm}
\begin{eqnarray}
f_{\Omega_i}(\omega) =
      \displaystyle \omega^{-\frac{2+\alpha}{\alpha}} \frac{\left[ \zeta \left(c_i \omega r_{\mathsf{net}}^{\alpha} \right) -\zeta \left( c_i \omega r_{\mathsf{ex}}^{\alpha} \right) \right] }{\alpha c_i^{2/\alpha} \left( r_{\mathsf{net}}^2 -r_{\mathsf{ex}}^2\right) }
\label{shad8}
\end{eqnarray}
for $0 \leq \omega \leq \infty$, and zero elsewhere,
where
\begin{eqnarray}
\zeta( z )= \mbox{erf} \left(  \frac{\sigma_s^2 \ln^2(10)-50 \alpha \ln\left( z \right)}{5 \sqrt{2} \alpha \sigma_s \ln(10)}\right) e^{ \frac{\sigma_s^2 \ln^2(10)}{50 \alpha^2}}
\label{zeta}
\label{zeta}
\end{eqnarray}

Because $||X_0||$ is deterministic, $\Omega_0 = 10^{\xi_0/10} ||X_0||^{-\alpha}$ is a log-normal variable with pdf
\begin{eqnarray}
f_{\Omega_0}(\omega)
=
\frac{10 \left(2 \pi \sigma_s^2\right)^{-\frac{1}{2}}}{\ln(10) \omega}  \exp\left \{ {-\frac{10^2 \log_{10}^2 \left( ||X_0||^{\alpha} \omega \right) }{2 \sigma_s^2}}\right \}
\label{shad02}
\end{eqnarray}
for $0 \leq \omega \leq \infty$, and zero elsewhere.

By substituting (\ref{shad02}) and (\ref{shad8}) into (\ref{cdf_M_rho}) and using the definition of $\beta_0$, the complementary outage probability is found to be
%\begin{eqnarray}
%\bar{F}_{\mathsf Z}(z) =   \int  f_{\Omega_0}(\omega) \bar{F}_{\mathsf Z}(z \big| \omega ) d\omega
%\end{eqnarray}
%and then
(\ref{shad05}), given at the bottom of the page,
where
\setcounter{equation}{22}
\begin{eqnarray}
\varphi(\omega)
\hspace{-0.2cm}
& = &
\zeta \left( c_i \omega r_{\mathsf{net}}^{\alpha}  \right) -\zeta \left( c_i \omega r_{\mathsf{ex}}^{\alpha}  \right) \\
\kappa(x,t)
\hspace{-0.2cm}
& = &
\hspace{-0.2cm}
\frac{
p_{i}\Gamma(\ell_{i}+m_{i})
x^{-\left( \frac{2+\alpha}{\alpha} \right) }
\left(  \frac{x}{m_{i}  }\right)  ^{\ell_{i}}
}{
\alpha c_i^{2/\alpha} (r_{\mathsf{net}}^2 - r_{\mathsf{ex}}^2)
(\ell_{i}! )
\Gamma(m_{i})
\left(  \frac{ x t}{m_{i} }+1\right)^{(m_{i}+\ell_{i})}
}. \nonumber \\
\label{Psi}
\end{eqnarray}

In order to compute the outage probability, the integral inside the product in  (\ref{shad05}) can be evaluated numerically by Simpson's method, which provides a good tradeoff between accuracy and speed, while the second integral can be evaluated through Monte Carlo simulation.

\begin{figure}[t]
\centering
%\vspace{-0.3cm}
\hspace{-0.5cm}
\includegraphics[width=9.25cm]{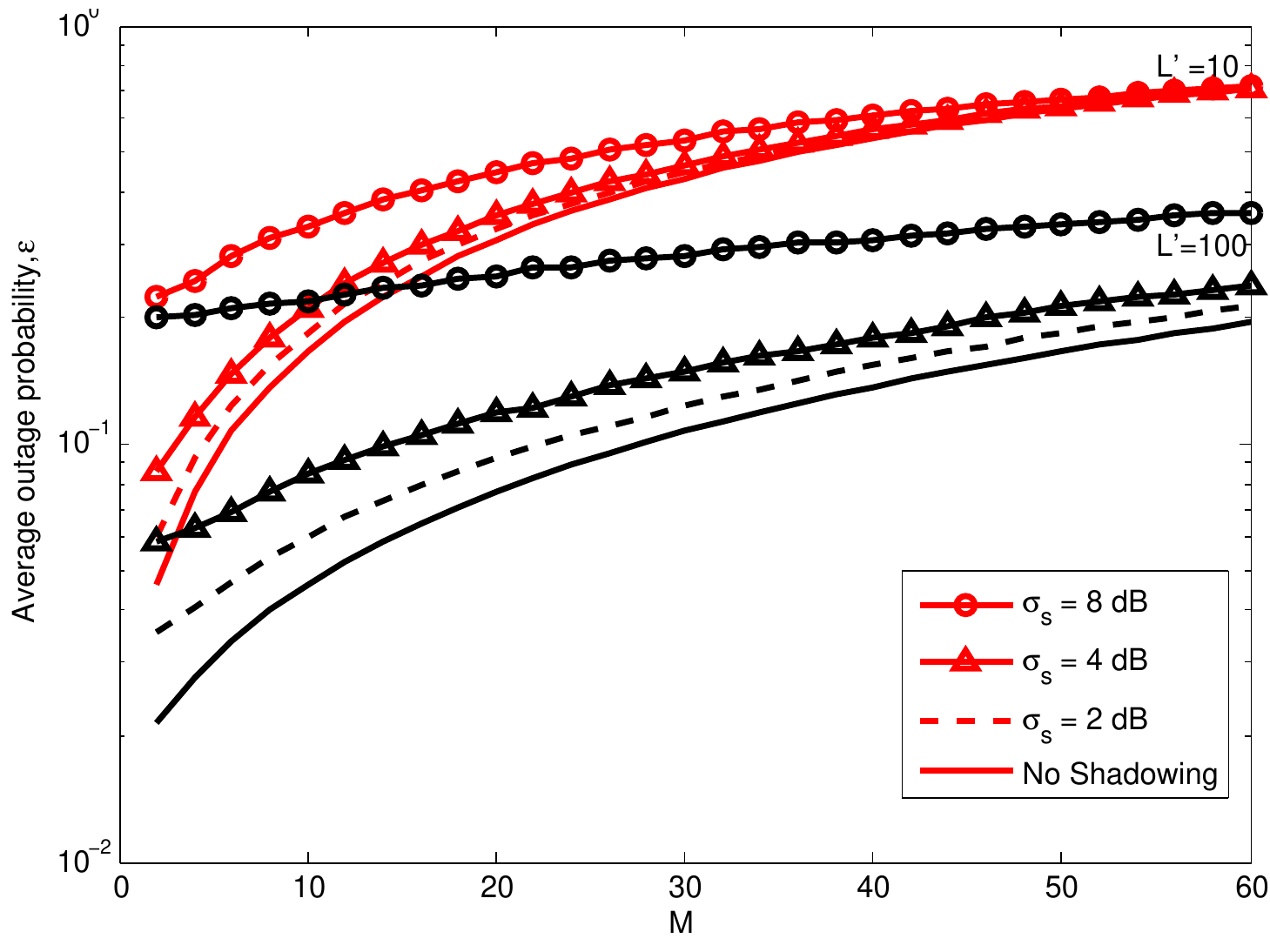}
\vspace{-0.7cm}
\caption{Average outage probability as a function of the number of interferers $M$ for two values $L'$.  For each $L'$, curves are shown for the case of no shadowing, and for shadowing with three values of $\sigma_s$.    \label{Figure:2} }
\vspace{-0.5cm}
\end{figure}

{\bf Example \#3.} The average outage probability is shown as a function of $M$ for two $L'$ in Fig. \ref{Figure:2}.  The $M$ interferers are uniformly located within the annulus bounded by $r_{\mathsf{ex}}=0.25$ and $r_{\mathsf{net}}=4$.  The results assume a path-loss exponent $\alpha = 3$, a common transmit power $P_i=P_0$, and that the source is at unit distance.  A mixed-fading model is assumed, i.e. $m_0 = 4$ and $m_i=1$ for $i \geq 1$, the SNR is $\Gamma = 10$ dB, and the SINR threshold is $\beta = 3.7$ dB.  For each value of $L'$, four curves are shown, one without shadowing and three with shadowing. To model log-normal shadowing, each $\xi_i$ drawn from a zero-mean Gaussian distribution with standard deviation $\sigma_s$ that varies among the set $\sigma_s = \{ 2, 4, 8\}$ dB.  From Fig. \ref{Figure:2}, it is observed that the outage probability degrades with increasing $M$ and decreasing $L'$, and that shadowing increases the average $\epsilon$ increasing $\sigma_s$ for the given parameters.

\section{Transmission Capacity}\label{Section:TC}
Often, networks are constrained to ensure that the outage probability $\epsilon$ does not exceed a maximum outage probability $\zeta$ $\in$ $\left[ 0,1 \right]$; i.e., $\epsilon(\lambda) \leq \zeta$ with the dependence of $\epsilon$ on the density of transmissions $\lambda$ made explicit.  Under such a constraint, the maximum density of transmissions is of interest, which is quantified by the {\em transmission capacity} (TC) \cite{weber:2005,weber:2010}.  With outage constraint $\zeta$, the TC is
\begin{eqnarray}
\tau_c\left(\zeta \right)
 & = &
\epsilon^{-1}(\zeta)(1-\zeta)
\label{TC_definition}
\end{eqnarray}
where $\epsilon^{-1}(\zeta)$ is the density of the underlying process whose spatially averaged outage probability satisfies the constraint $\epsilon(\lambda) \leq \zeta$ with equality,
%\footnote{Since $\epsilon$ is a monotonically increasing function of $\lambda$, the TC is maximized when the constraint $\epsilon \leq \zeta$ is met with equality.}
and $(1-\zeta)$ ensures that only successful transmissions are counted.   The TC %represents the spatial spectral efficiency; i.e. the rate of successful data transmission per unit area.  With appropriately normalized variables, the TC can assume units of bits-per-second per Hz per $m^2$ (bps/Hz/$m^2$).
is a measure of the spatial intensity of transmissions and has the units of number of (successful) transmissions per unit area.

As originally defined in \cite{weber:2005}, the transmission capacity is a function of the SINR threshold $\beta$ and is found without making any assumptions about the existence of any particular type of modulation or channel coding.  In practice, the SINR threshold is a function of the modulation and coding that is used.   Let $C( \gamma )$ denote the maximum achievable rate that can be supported by the chosen modulation at an instantaneous SINR of $\gamma$ assuming equally likely input symbols; i.e., it is the modulation-constrained capacity or symmetric-information rate.  If a rate $R$ code is used, then an outage will occur when $C(\gamma) \leq R$.   Since $C(\gamma)$ is monotonic, it follows that $\beta$ is the value for which $C(\beta)=R$, and therefore we can write $\beta = C^{-1}(R)$.  If it is assumed that each codeword is transmitted using one hop, then the outage probability with SINR threshold $\beta = C^{-1}(R)$ provides a good prediction of the codeword error rate.

Frequency-hopping systems often use noncoherent CPFSK modulation \cite{cheng:ciss2007,torrieri:2011}.  The maximum achievable rate of noncoherent CPFSK is given in \cite{cheng:ciss2007} for various modulation indices $h$, where it is called the {\em symmetric information rate}. In particular, Fig. 1 of \cite{cheng:ciss2007} shows the symmetric information rate of binary CPFSK as a function of $\gamma$ for various $h$.  To emphasize the dependence of the capacity on $h$, we use $C(h,\gamma)$ in the sequel to denote the rate of CPFSK with modulation index $h$.  For any value of $h$, the value of the SINR threshold $\beta$ can be found from the corresponding curve by finding the value of $\gamma$ for which $C(h,\gamma)=R$.  For instance, when $R=1/2$ and $h=1$, the required $\beta = 3.7$ dB.  In \cite{torrieri:2008}, it was found that in practice, and over a wide range of code rates, turbo-coded noncoherent CPFSK is consistently about 1 dB away from the corresponding modulation-constrained capacity limit.  Thus, the $\beta$ required in practice will generally be higher than the value obtained from the capacity limit by a small margin.  For instance, if a 1 dB margin is used, then the SINR threshold for noncoherent binary CPFSK with $R=1/2$ and $h=1$ should be set to $\beta = 4.7$ dB.
% CORRECTION ABOVE: Replaced $\beta_{min}$ with capacity limit.

When accounting for modulation and coding, the maximum data transmission rate is determined by the bandwidth $B/L$ of a frequency channel, the spectral efficiency of the modulation, and the code rate.  Let $\eta$ be the spectral efficiency of the modulation, given in symbols per second per Hz, and defined by the symbol rate divided by the 99 percent-power bandwidth of the modulation\footnote[1]{Percent-power bandwidths other than 99 can be used, but will influence the amount of adjacent-channel interference.}.  The spectral efficiency of CPFSK can be found by numerically integrating the normalized power-spectral densities given in \cite{torrieri:2011}, or since we assume many symbols per hop, by Equation (3.4-61) of \cite{proakis:2008} and then inverting the result.  To emphasize the dependence of $\eta$ on $h$, we denote the spectral efficiency of CPFSK as $\eta(h)$ in the sequel.  When combined with a rate-$R$ code, the spectral efficiency of CPFSK becomes $R \eta(h)$ (information) bits per second per Hz, where $R$ is the ratio of information bits to code symbols.  The data rate supported by the channel is $R \eta(h) B/L$ bits per second.  The average data rate, or throughput, must account for the duty factor $D$ and only count correct transmissions.  Hence, the throughput is
\begin{eqnarray}
   T
   & = &
   \frac {   R \eta(h) B D (1-\epsilon) }{L}
   =
   \frac {   R \eta(h) B (1-\epsilon) }{L'}.
\end{eqnarray}

The {\em modulation-constrained} transmission capacity is the throughput multiplied by the node density,
\begin{eqnarray}
   \tau (\lambda)
   & = &
   \mathcal \lambda T = \frac{\lambda R \eta(h) B (1-\epsilon) }{L'}.\label{Equation:TC}
\end{eqnarray}
where $\lambda = M/(\pi(r_{\mathsf{net}}^2-r_{\mathsf{ex}}^2))$ is the number of interferers per unit area.  In contrast with (\ref{TC_definition}), this form of transmission capacity explicitly takes into account the code rate $R$, as well as the spectral efficiency of the modulation $\eta(h)$.  It furthermore accounts for the hopping bandwidth $B/L'$.  Rather than constraining outage probability, $\tau( \lambda )$ and the outage probability vary with the node density $\lambda$.  Since it accounts for the actual system bandwidth $B$, (\ref{Equation:TC}) assumes units of $bps/m^2$.  By dividing by bandwidth, the {\em normalized} modulation-constrained transmission capacity
\begin{eqnarray}
  \tau'(\lambda)
  & = &
  \frac{\tau}{B}
  =
  \frac{\lambda R \eta(h) (1-\epsilon) }{L'}.\label{Equation:TCnorm}
\end{eqnarray}
takes on units of $bps/Hz/m^2$.  However, unlike (\ref{TC_definition}), $\tau'(\lambda)$ is in terms of {\em information} bits rather than {\em channel} bits.

%Note that (\ref{Equation:TC}) requires that the outage probability be independent of the location of the reference receiver, which is true when the network is assumed to be infinite and uniform.  Otherwise, the throughput needs to be computed at each possible location, and the spatial average taken.

\section{Network Optimization}\label{Section:Optimization}
For a particular spatial distribution and channel model, there will be a set of $(L',R,h)$ that maximizes the normalized TC $\tau'(\lambda)$.  Finding the optimal set of parameters is challenging because the search space is large.  However, in \cite{talarico:2012}, we found that the optimization surface is convex in the case of unshadowed Rayleigh fading, and further experiments have confirmed that the surface is still convex in the presence of shadowing and Nakagami fading.

When the search space is convex, the optimal set of $(L',R,h)$ can be found as follows:
 \begin{enumerate}
   \item For each $L'$, $R$, and $h$ select the endpoints for the search interval. Initially, the endpoints are widely separated to ensure that the optimal point lies in the search interval.
   \item Compute $\tau'(\lambda)$ for the endpoints and the midpoint of the interval of one of the parameters in the set $(L',R,h)$. \label{step2}
   \item For the same parameter of the set $(L',R,h)$ used in step \ref{step2}, move the midpoint of the interval towards the endpoint that gives higher $\tau'(\lambda)$.  Reduce the size of the interval by moving the endpoints closer.
    \label{step3}
   \item Repeat step \ref{step2} and \ref{step3} recursively for all the parameter in the set $(L',R,h)$, until the following conditions are both satisfied:
       \begin{enumerate}
       \item The midpoints of all three intervals are the same as in the previous iteration;
       \item The difference between the endpoints and the midpoint of the interval is equal to a given tolerance.
       \end{enumerate}
\end{enumerate}

Optimization results were obtained for $M=50$ interferers at a fixed SNR of $\Gamma = 10$ dB.  The networks had an inner radius of $r_{\mathsf{ex}} = 0.25$ and one of two maximum radii: $r_{\mathsf{net}} = 2$, which creates a dense network, or $r_{\mathsf{net}} = 4$, which creates a sparser network.  Three sets of fading coefficients were considered: $m_i = 1$ for all transmitters (universal Rayleigh fading),  $m_i=4$ for all transmitters (universal Nakagami fading), and a combination of $m_0 = 4$ for the source and $m_i = 1$ for the interferers (mixed fading).  Unshadowed ($\sigma_s = 0$ dB) and shadowed ($\sigma_s = 8$ dB) environments were examined.

The results of the optimization are shown in Table \ref{maintable}.  For each set of parameters, the $(L',R,h)$ that maximize the TC are listed, along with the corresponding normalized TC $\tau_{opt}'$.  In addition to showing the TC when using the optimal parameters, the normalized TC $\tau_{sub}'$ is shown for a typical choice of parameters: $(L',R,h) = (200,1/2,1)$.

When $R$ is related to $\beta$ through the modulation-constrained AWGN capacity, the optimization assumes Gaussian interference and capacity-achieving codes.  However, as mentioned in Section \ref{Section:TC}, actual systems require a threshold that is higher than this ideal value by some margin.  The column labeled $\tau'_1$ gives the normalized TC when a 1 dB margin is used with the listed values of $(L',R,h)$.   A modest loss in TC is observed when this margin is used.

\begin{table}
  \centering
  \caption{Results of the Optimization for $M=50$ interferers.  The normalized TC $\tau'$ is in units of bps/kHz-$m^2$. \label{maintable}}
   \vspace{-0.15cm}
  \begin{tabular}{|c|c|c|c|c|c|c|c|c|c|}
  \hline
  $r_{\mathsf{net}}$ & $\sigma_s$ & $m_0$ & $m_i$ & $L'$ & $R$ & $h$ & $\tau'_{opt} $ & $\tau'_{1}$ & $\tau'_{sub}$ \\
  \hline
  2         & 0            &  1    &   1   &  32 & 0.62  & 0.59  & 15.90 &13.57 & 3.34  \\
  \cline{3-10}
            &              &  4    &   4   &  42 & 0.66& 0.59 & 17.37 & 14.67  & 4.12  \\
  \cline{3-10}
            &              &  4    &   1   & 36 & 0.65& 0.59 & 20.15 & 16.96 & 4.19 \\
  \cline{2-10}
            & 8            &  1    &   1   & 23 & 0.72& 0.59 & 19.39 & 16.68 & 3.00 \\
  \cline{3-10}
            &              &  4    &   4   & 28 & 0.76& 0.59 & 19.74 & 16.98 & 3.43  \\
  \cline{3-10}
            &              &  4    &   1   &  24 & 0.68& 0.59 & 22.15 & 19.23 & 3.46  \\
  \hline
  4         & 0            &  1    &   1   &  12 & 0.54 & 0.59 & 9.83 & 7.98 & 0.90 \\
  \cline{3-10}
            &              &  4    &   4   &  15 & 0.50 & 0.59 & 10.83 & 8.63 & 1.13 \\
  \cline{3-10}
            &              &  4    &   1   &   13 & 0.50 & 0.59 & 12.03  & 9.57 & 1.13  \\
  \cline{2-10}
            & 8            &  1    &   1   &    9 & 0.66 & 0.59 & 10.62  & 8.94 & 0.78 \\
  \cline{3-10}
            &              &  4    &   4   &   10 & 0.62 & 0.59 & 11.05   & 9.10 & 0.91 \\
  \cline{3-10}
            &              &  4    &   1   &    9 & 0.65 & 0.59 & 12.35  & 10.41 & 0.91 \\
  \hline
  \end{tabular}
  \vspace{-0.55cm}
\end{table}

The results shown in Table \ref{maintable} highlight the importance of parameter optimization.  The TC is improved by a factor of 5-10 by selecting optimal, rather than arbitrary parameters.  The influence of the fading distribution can be observed.  Performance is worst under the assumption that all channels undergo Rayleigh fading, but this is a pessimistic assumption.  Performance improves when all channels have a common $m_i=4$, but it is unrealistic to assume that all interferers are in the line-of-sight.  Performance is further improved when the interferers undergo Rayleigh fading while the source undergoes better than Rayleigh fading ($m_0=4$).

The optimal modulation index is $h=0.59$ for all network configurations, which coincides with the optimal value found in \cite{cheng:ciss2007} for  Rayleigh fading.  The optimal $L'$ is larger for denser networks and when the signals become more line-of-sight.  The denser network requires a higher $R$ than the sparser network.  At first inspection, the last result may seem counter-intuitive. One would ordinarily expect a denser network to require the error-protection of a lower-rate code.  However, the sparser network uses a lower value of $L'$, and hence, has a higher probability of collision $p_i=1/L'$.  The lower-rate code used by the sparse network helps to offset this higher collision probability.

Table \ref{maintable} shows that performance actually improves in the presence of shadowing.  This is because with shadowing, the outage probability is less sensitive to the choice of code rate $R$ and the number of hopping channels $L'$.   While the outage probability is worse with shadowing than without for a wide range of $R$ and $L'$, when the value of $R$ is sufficiently high and/or the value of $L'$ sufficiently low, the outage probability is improved with shadowing.  As can be seen in Table \ref{maintable}, the optimal values of $R$ and $L'$ are larger and smaller, respectively, with shadowing than without.  As indicated by (\ref{Equation:TC}), increasing $R$ or decreasing $L'$ will increase the transmission capacity for a given outage probability and modulation index.

\balance

\section{Conclusion} \label{Section:Conclusion}
Frequency-hopping ad hoc networks play an important role in modern communications.  When used with coded CPFSK, which is the typical choice of modulation, the performance depends on the modulation index, the code rate, and the number of frequency-hopping channels.  These parameters are often chosen arbitrarily.  The procedure disclosed in this paper enables the optimization of the parameters for a fixed network with a fixed number of users in the presence of shadowing and Nakagami fading.  The key innovation facilitating the optimization is a new closed-form expression for the outage probability in the presence of Nakagami fading.

%The optimization can be performed with respect to a particular spatial distribution by generating a representative sampling of the distribution and averaging the outage probability with respect to the drawn distribution.  This approach allows any spatial model with a fixed number of users to be considered.  The methodology described in this paper can easily be extended to accommodate a variable number of users.  This can be done using one of two approaches: (1) rather than drawing networks with fixed $M$, networks may be drawn that have a random $M$, or (2) the outage probability can be first computed for each possible value of $M$ and then the average taken with respect to the distribution of $M$.

The results presented in this paper are just a sample of what is possible using this methodology.  For instance, other types of modulation and reception could be considered, such as nonbinary CPFSK or multi-symbol noncoherenet reception \cite{valenti:2010}.   The influence of the spatial model may be studied, as can the use of directional antennas.  The role of adjacent-channel interference can be studied by considering the effect of spectral splatter.  The influence of receiver position can be studied by allowing it to move radially from the center of the network to the outer perimeter.

% Such an analysis may suggest that using the 99\%-power bandwidth to space adjacent channels may not be optimal.

\bibliographystyle{ieeetr}
\bibliography{icc2012refs}

\end{document}